\documentclass[12pt,preprint]{aastex}
\usepackage{graphicx}
\shortauthors{Allers et. al.}
\shorttitle{2M2234+4041AB}

\begin{document}

\newcommand{\Ks}{\mbox{$K_S$}}
\newcommand{\Lp}{\mbox{$L^{\prime}$}}
\newcommand{\sciencebin}{\hbox{2M2234+4041AB}}
\newcommand{\sciencebina}{\hbox{2M2234+4041A}}
\newcommand{\sciencebinb}{\hbox{2M2234+4041B}}
\newcommand{\degs}{\mbox{$^{\circ}$}}
\newcommand{\etal}{et al.}
\newcommand{\eg}{e.g.}
\newcommand{\ie}{i.e.}

\title{2MASS~22344161+4041387AB: A Wide, Young, Accreting, Low-mass Binary in the LkH$\alpha$233 Group\footnote{Some of the data presented herein were obtained at the W.M. Keck Observatory, which is operated as a scientific partnership among the California Institute of Technology, the University of California and the National Aeronautics and Space Administration. The Observatory was made possible by the generous financial support of the W.M. Keck Foundation.}}
\author{K.~N.~Allers\footnote{Visiting Astronomer at the Infrared Telescope Facility, which is operated by the University of Hawaii under Cooperative Agreement no. NCC 5-538 with the National Aeronautics and Space Administration, Office of Space Science, Planetary Astronomy Program.}, Michael~C.~Liu\footnote{Alfred P. Sloan Research Fellow}} 
\affil{Institute for Astronomy, University of Hawaii, 
2680 Woodlawn Drive, Honolulu, HI 96822}
\author{Evgenya~Shkolnik\footnote{Carnegie Fellow}}
\affil{Department of Terrestrial Magnetism,
Carnegie Institute of Washington,
5241 Broad Branch Road, NW, Washington, DC 20015}
\author{Michael~C.~Cushing\footnotemark[2], Trent~J. Dupuy, Geoffrey~S.~Mathews\footnotemark[2]} 
\affil{Institute for Astronomy, University of Hawaii, 
2680 Woodlawn Drive, Honolulu, HI 96822}
\author{I. Neill Reid}
\affil{Space Telescope Science Institute, 3700 San Martin Drive, Baltimore, MD 21218}
\author{Kelle~L.~Cruz\footnote{Spitzer Fellow}} 
\affil{Department of Astronomy, California Institute of Technology, Pasadena, CA 91125}
\author{W.~D.~Vacca\footnotemark[2]}
\affil{SOFIA-USRA, NASA Ames Research Center, MS N211-3, Moffett Field, CA 94035}

\begin{abstract}
We report the discovery of a young, 0.16\arcsec\ binary, \sciencebin, found as the result of a Keck laser guide star adaptive optics imaging survey of young field ultracool dwarfs.  Spatially resolved near-infrared photometry and spectroscopy indicate that the luminosity and temperature ratios of the system are near unity.  From optical and near-infrared spectroscopy, we determine a composite spectral type of M6 for the system.  Gravity-sensitive spectral features in the spectra of \sciencebin\ are best matched to those of young objects ($\sim$1~Myr old).   A comparison of the T$_{eff}$ and age of \sciencebin\ to evolutionary models indicates the mass of each component is 0.10$^{+0.075}_{-0.04}$ M$_\odot$.  Emission lines of H$\alpha$ in the composite optical spectrum of the system and Br$\gamma$ in spatially resolved near-IR spectra of the two components indicate that the system is actively accreting.   Both components of the system have IR excesses, indicating that they both harbor circumstellar disks.  Though \sciencebin\ was originally identified as a young field dwarf, it lies 1.5\arcmin\ from the well-studied Herbig Ae/Be star, LkH$\alpha$233.  The distance to LkH$\alpha$233 is typically assumed to be 880~pc. It is unlikely \sciencebin\ could be this distant, as it would then be more luminous than any known Taurus objects of similar spectral type.   We re-evaluate the distance to the LkH$\alpha$233 group and find a value of 325$^{+72}_{-50}$~pc, based on the \emph{Hipparcos} distance to a nearby B3-type group member (HD~213976).  \sciencebin\ is the first low-mass star to be potentially associated with the LkH$\alpha$233 group.  At a distance of 325~pc, its projected physical separation is 51~AU, making it one of a growing number of wide, low-mass binaries found in young star-forming regions.
\end{abstract}


\section{Introduction}

The binary fractions, mass ratios and separations of very low-mass (VLM; M$_\star \lesssim$0.1~M$_\odot$) binaries can provide critical tests to theories of VLM star and brown dwarf formation.  
In addition to the same mechanism that forms low-mass stars, there are a number of suggested processes for forming brown dwarfs \citep{whitworth07}. These include gravitational instabilities in massive circumstellar disks \citep[e.g.][]{stamatellos08}, turbulent or gravitationally enhanced fragmentation \citep[e.g.][]{padoan04,bonnell08}, ejection from protostellar embryos \citep[e.g.][]{bate05,clarke01}, and photoevaporation of protostellar cores \citep{whitworth04}.  These various formation mechanisms proposed for brown dwarfs are expected to result in differing binary fractions, mass ratios, and separations.

To date, imaging surveys for VLM binaries have focused on field ($\gtrsim$1~Gyr old) objects \citep[e.g.][]{burgasser06,goldman08,reid06,siegler05, liu09}, and have found much lower binary fractions (7--15\% reported by \citet{burgasser07};  20\% if corrected for completeness \citep{allen07}) than found for solar-type field stars \citep[44\%,][]{duquennoy91}.    The detected field VLM binaries also tend to be tightly bound and in nearly equal mass systems, whereas higher mass field binaries have a much wider range of separations and mass ratios. 
It is difficult, however, to robustly test models of brown dwarf formation with field binary statistics for a number of reasons.  First, wide VLM binaries could be dynamically disrupted by the time the systems reach the age of the field population \citep{close07}.   Additionally, the intrinsic faintness and large changes in luminosity with mass for field objects \citep[e.g. ][]{baraffe03} mean that low-mass ratio systems were unlikely to be detected by existing high resolution imaging surveys.  Lastly, the field population is comprised of objects that were formed in a variety of star-forming regions with varying initial conditions, so comparison to model calculations for brown dwarfs forming in a single molecular cloud may be inappropriate.

To provide a better test of brown dwarf and star formation models, one would like to compare the binary properties of young, VLM objects to model predictions.
The statistics for young, low-mass binaries are still uncertain.  \citet{ahmic07} combined their survey with results from Kraus et al. (2005, 2006) and \citet{konopacky07} and report a combined binary fraction of 7$^{+4}_{-3}$\% for 72 young, low-mass stars and brown dwarfs.  This binary fraction is similar to the binary fraction for field VLM binaries \citep[7--15\%,][]{burgasser07}.   However, because the nearest star-forming regions are $\sim$125~pc away, only wide binaries can be detected by current imaging surveys.  The \citet{ahmic07} binary fraction is valid only for young VLM binaries with separations greater than 8~AU, whereas field VLM binaries tend to have tighter separations \citep{burgasser07}.  Thus, the actual young binary fraction would be significantly higher than the fraction found for field VLM binaries, if surveys for young and field VLM binaries reached the same physical separation limits.  Young star-forming regions also show an interesting population of very wide separation ($\gtrsim$100~AU), loosely bound VLM binaries \citep{bejar08, bouy06, close07, luhman04a}. The fate of these wide binaries is unknown, as it is unlikely that they could survive ejection or dispersal from their natal clouds, given their low binding energies. 

To probe the multiplicity properties of young VLM objects at tight separations and look for wide systems in the field that have not yet dissolved, we are carrying out a Keck laser guide star adaptive optics (LGS~AO) survey to image young ($\lesssim$100~Myr), field VLM objects.  This paper presents the first results of this program.  This survey is a part of our larger, ongoing effort using LGS~AO to study the multiplicity of VLM objects and determine their fundamental properties \citep[e.g.][]{dupuy08,liu08,liu06}.

Young field objects are usually found in three ways: as companions to known young field stars \citep{rebolo98}, via association with known young moving groups \citep[e.g.][]{gizis02}, or serendipitously as a part of searches for older field dwarfs \citep[e.g.][] {kirkpatrick08}.  2MASS~J22344161+4041387 (hereinafter \sciencebin) was identified by \citet{cruz03} as a candidate young M6-type object.  They classified the object as young on the basis of weak CaH (6750--7050 \AA) and KI doublet (7665 and 7699 \AA) absorption and strong H$\alpha$ emission seen in its low-resolution optical spectrum.   

Using Keck LGS~AO, we have discovered \sciencebin\ to be a binary system.  In this paper, we present multi-wavelength imaging and spectroscopy used to determine the properties of \sciencebin.

\section{Observations}

\subsection{Keck LGS~AO/NIRC2 Imaging}
\begin{figure}
\vskip -8in
\vskip 7.5in
\hskip 2.1in
\centerline{\includegraphics[width=6in,angle=90]{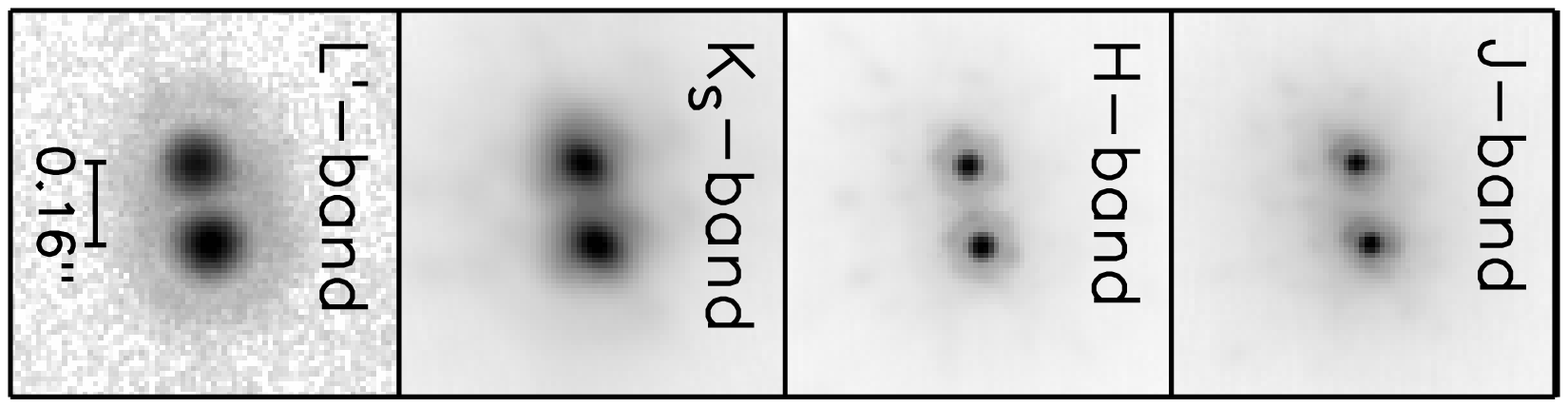}}
\vskip -6.02in
\hskip 4.2in
\centerline{\includegraphics[width=6in,angle=90]{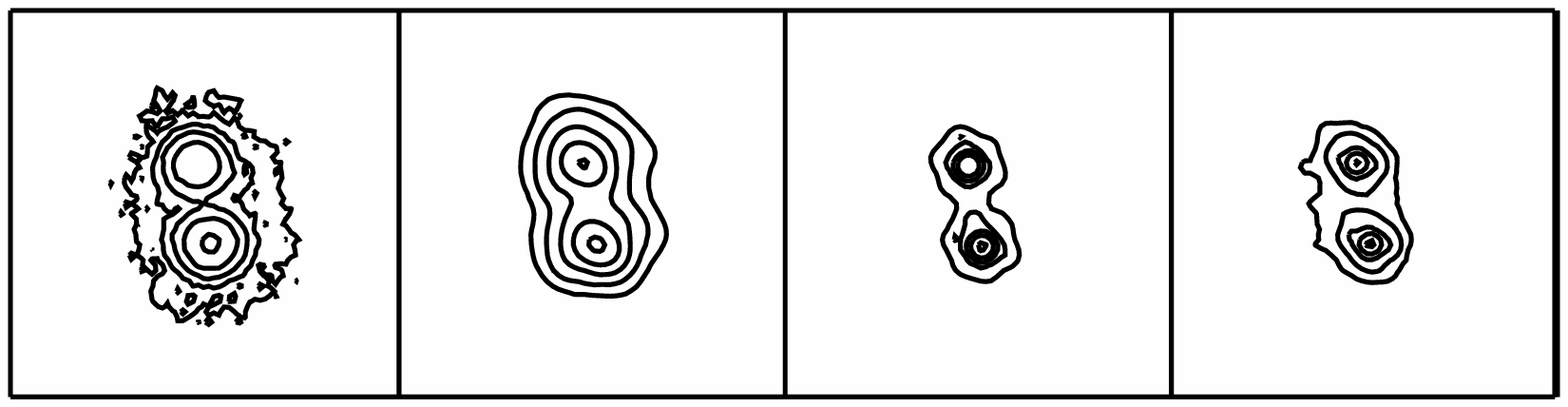}}
\vskip 2ex
\caption{\label{nircims} $JH\Ks\Lp$-band imaging of \sciencebin\ from Keck LGS
  AO.  North is up and east is left.  Each image is 0.75\arcsec\ on a
  side. The greyscale image uses a square-root intensity stretch.
  Contours are drawn from 90\%, 45\%, 22.5\%, 11.2\% and 5.6\% of
  the peak value in each bandpass.}
\end{figure}

We imaged \sciencebin\ on three epochs from 2006--2008 using the laser
guide star adaptive optics (LGS AO) system \citep{2006PASP..118..297W,
  2006PASP..118..310V} of the 10-meter Keck II Telescope on Mauna Kea,
Hawaii.  We used the facility IR camera NIRC2 with its narrow
field-of-view camera, which produces a $10.2\arcsec \times
10.2\arcsec$ field of view, and the Mauna Kea Observatories $J, H, \Ks$ 
and \Lp\ filters \citep{simons02,tokunaga02}.  
Images of \sciencebin\ are presented in Figure \ref{nircims}.  
Setup time for the telescope to slew to the science target and for the
LGS AO system to be fully operational varied from 4--21~minutes on the
observing runs.  The LGS brightness was equivalent to a
$V\approx9.4-10.1$~mag star, as measured by the flux incident on the
AO wavefront sensor.  The LGS provided the wavefront reference
source for AO correction, with the exception of tip-tilt motion.  Tip-tilt aberrations and quasi-static changes in
the image of the LGS as seen by the wavefront sensor were measured
contemporaneously with a second, lower-bandwidth wavefront sensor
monitoring \sciencebin\ itself (R=16.8;
\citealp{2003AJ....125..984M}).  The sodium laser beam was pointed at
the center of the NIRC2 field-of-view for all observations.

For the $JH\Ks$ filters, we obtained a series of dithered images,
offsetting the telescope by a few arc-seconds. 
The resulting images were reduced in the 
standard fashion (dome flat-fielded, median sky-subtracted, registered and stacked).
The \Lp-band data were observed and reduced in a similar fashion, except
that the flat fields were constructed from the science frames themselves
and sky
subtraction was done in a pairwise fashion using consecutive frames.

To measure the flux ratios and relative positions of the binary's two
components, we used an analytic model of the point spread function
(PSF) as the sum of three elliptical gaussians.  For the individual
images obtained with each filter, we fitted for the flux ratio,
separation, and position angle of the binary.  The averages of the
results were adopted as the final measurements and the standard
deviations as the errors.  The relative astrometry was corrected for
instrumental optical distortion based on analysis by
B. Cameron (priv.\ comm.) of images of a precisely machined pinhole
grid located at the first focal plane of NIRC2.  Since the binary
separation and the imaging dither steps are relatively small, the
effect of the distortion correction is minor.

In order to gauge the accuracy of our measurements, we created myriad
artificial binary stars from images of single stars, chosen to have
comparable Strehl and FWHM as the science data.  For each filter, our
fitting code was applied to artificial binaries with similar
separations and flux ratios as \sciencebin\ over a range of PAs.
These simulations showed that any systematic offsets in our fitting
code are very small, well below the random errors, and that the random
errors are accurate.  In cases where the RMS measurement errors from
the artificial binaries were larger than those from the \sciencebin\
measurements, we conservatively adopted the larger errors.  The
exception was the \Lp-band data, where the prominent first Airy ring
seen in the images coincides with the separation of the binary.
The results from the simulated \Lp-band binaries indicated a
significant ($1.5\sigma$) shift in the astrometry, so this was applied
to the raw measurements; the shift moves the \Lp-band astrometry into
better agreement with the $JH\Ks$ results.

To convert the instrumental measurements of the binary's separation
and PA into celestial units, we used a weighted average of the
calibration from \citet{2006ApJ...649..389P}, with a pixel scale of
$9.963\pm0.011$~mas/pixel and an orientation for the detector's
+y~axis of $0.13\pm0.07\degs$ east of north for NIRC2's narrow camera
optics.  These values agree well with Keck Observatory's notional
calibrations as well as the $9.963 \pm 0.005$~mas/pixel and $0.13 \pm
0.02$\degs\ reported by \citet{ghez08}.  Table~\ref{nirc2} presents
our final Keck LGS imaging measurements.


\subsection{IRTF/SpeX Near-IR Spectroscopy and \Lp\ Imaging}
\begin{figure}
\vskip 0.5in
\centerline{\includegraphics[width=6in,angle=90]{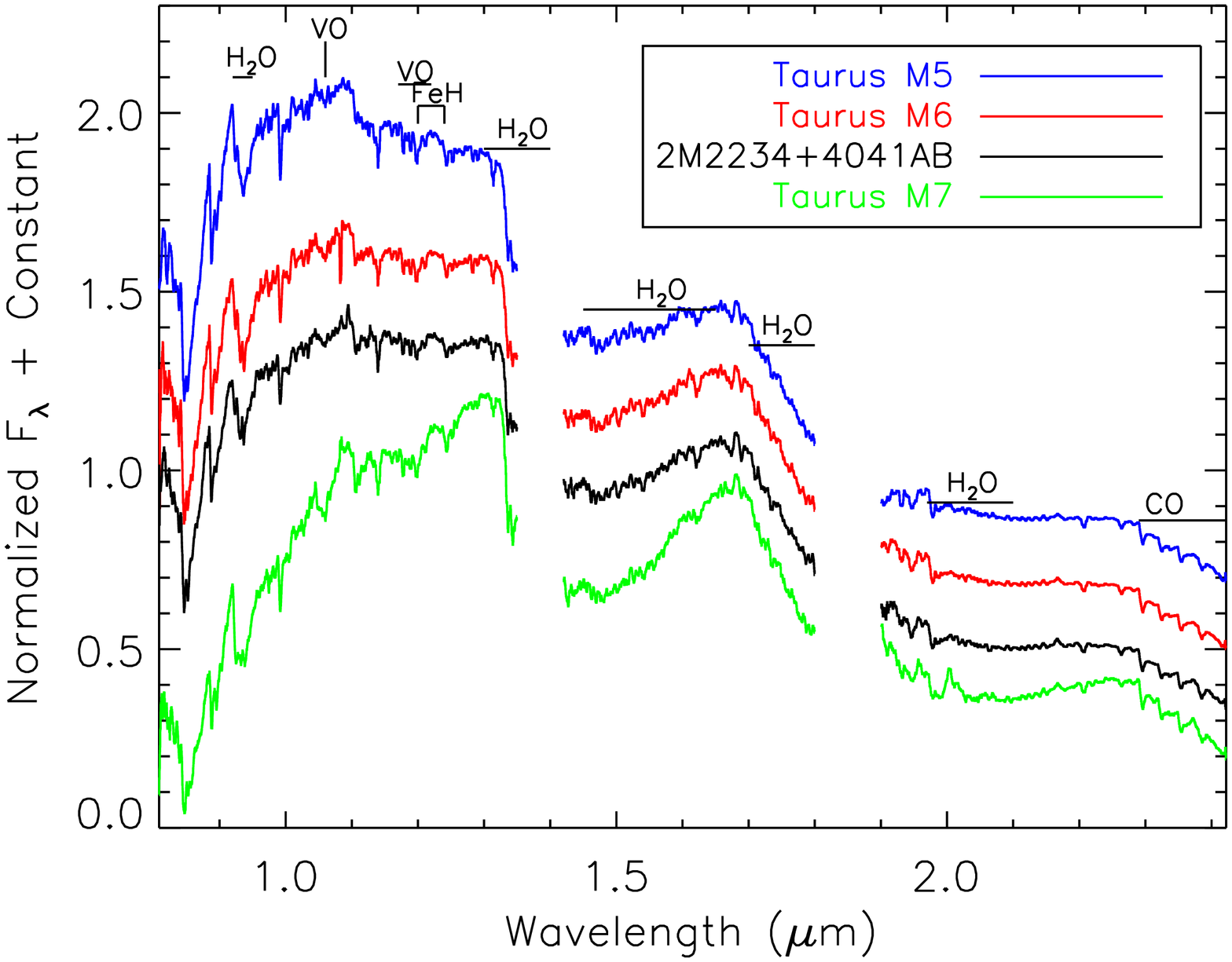}}
\vskip 2ex
\caption[The Near-Infrared Spectral Type of \sciencebin]
{\label{nirspt} Composite SpeX spectrum of \sciencebin\ (black) compared to young Taurus objects with optical spectral types of M5 (Haro~6-32, blue), M6 (CFHT-BD-Tau~18, red), and M7 (CFHT-BD-Tau~4, green).  The spectra of Taurus objects are smoothed to $\lambda/\Delta\lambda\sim500$ and dereddened \citep[using the reddening law of ][]{fitzpatrick} by A$_V$'s reported in \citet{guieu06} and \citet{luhman04b}.  Spectral type sensitive features are labeled.  The spectrum of \sciencebin\ (particularly its H$_2$O features) is best matched by a young Taurus M6.}
\end{figure}

Near-IR spectroscopy of \sciencebin\ was obtained on 2008 June 25 (UT) using the SpeX spectrograph \citep{rayner03} on the NASA Infrared Telescope Facility.  The seeing recorded by the IRTF was 0.6\arcsec. Six cycles of 180 second exposures were taken, nodding along the slit, for a total integration time of  36 minutes. The data were taken in SXD mode with a 0.3\arcsec\ slit (aligned with the parallactic angle) producing a 0.9--2.5~$\mu$m spectrum with a resolution ($\lambda/\Delta\lambda$) of $\sim$2000.  For telluric and flux correction, we observed a nearby A0V star, BD+39~4890 (V=9.47 mag) and obtained calibration frames (flats and arcs) prior to our observations of \sciencebin.  For comparison to our spectrum of \sciencebin, we obtained SpeX spectra of USco CTIO 75 \citep[M6,][]{ardila00} on 2008 Jun 25 and spectra of Haro~6-32 \citep[M5,][]{luhman04b}, CFHT-BD-Tau~18 \citep[M6,][]{guieu06}, and CFHT-BD-Tau~4 \citep[M7,][]{luhman04b} on 2005 October 22.  We used the same instrument setup as for our spectrum of \sciencebin, and obtained total integration times of 20 to 30 minutes per source.  The spectra were reduced using Spextool \citep{cushing04}, the facility reduction pipeline, which includes a correction for telluric absorption following the method described in \citet{vacca03}. Our SpeX spectra are presented in Figures \ref{nirspt} and \ref{spex}.

To obtain integrated-light \Lp\ photometry \citep[in the MKO system;][]{simons02,tokunaga02} of \sciencebin\ (Table \ref{binphot}), we obtained images on 2008 September 24 (UT) using the SpeX guider camera.  Conditions were photometric and the IRTF recorded a seeing of $\sim$0.6\arcsec.  We obtained a set of 10 nodded cycles of 300s exposures, nodding the telescope 7.7\arcsec\ in the north-south direction.  We repeated this process for 2 additional positions.  Between sets of \sciencebin\ images, we obtained images of the UKIRT faint standard FS2--27, using the same instrument and observing configuration.  To flat field our images, we used the median of dome flats for the J, H, K, and HK filters (kindly provided to us by John Rayner).  We subtracted each nodded pair, and median combined the resulting images at each position, for 6 final images of \sciencebin\ and 4 final images of FS2--27.    We obtained aperture photometry for both \sciencebin\ and FS2--27 using an aperture radius of 1.2\arcsec\ (10 pixels).  We calculated the mean and standard deviation of the fluxes measured at each position to determine our final photometry.  We measured an integrated light flux of $\Lp = 10.52\pm0.06$~mag for \sciencebin.

\subsection{Keck LGS~AO/OSIRIS Spectroscopy}
\begin{figure}
\vskip 0.5in
\centerline{\includegraphics[width=6in,angle=90]{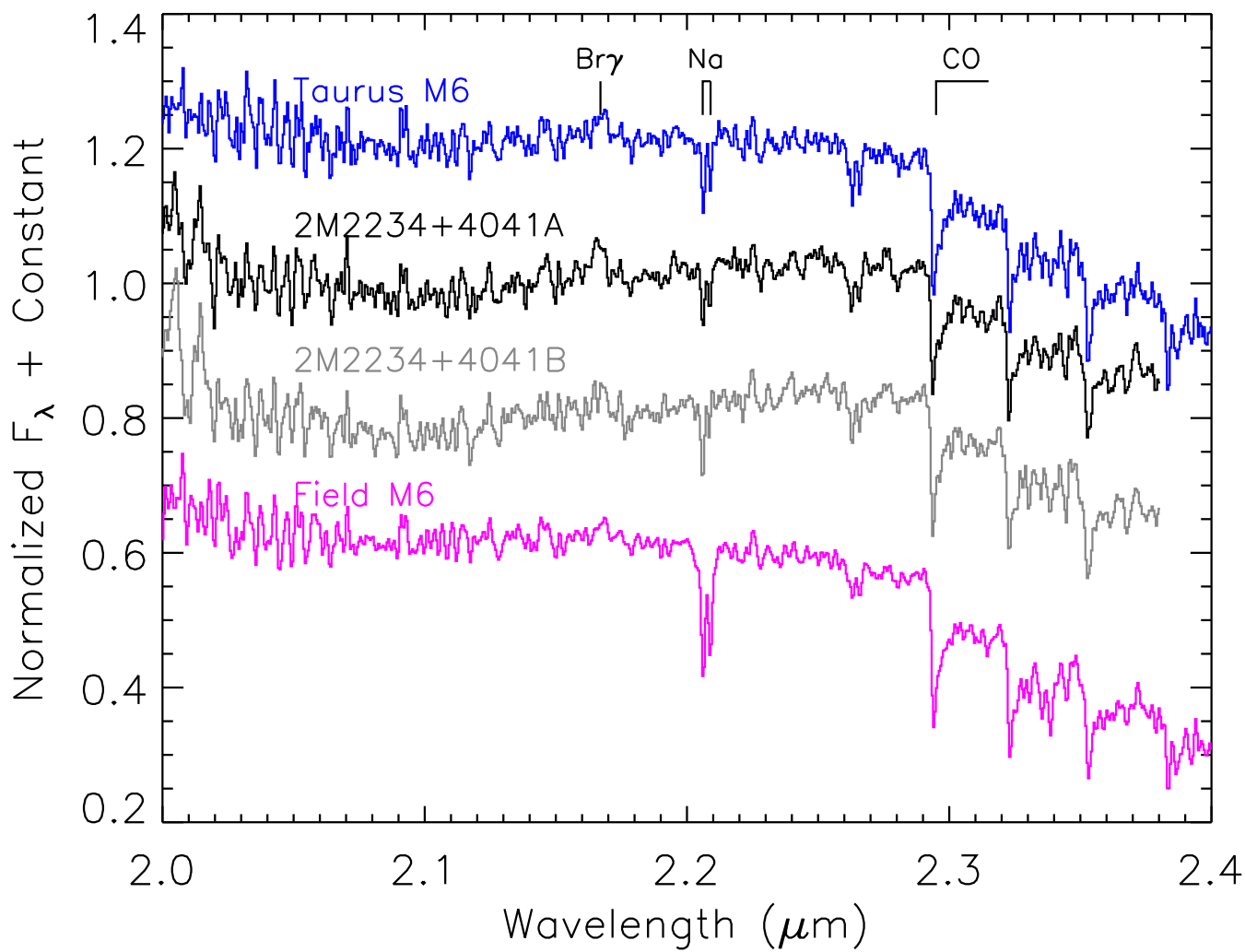}}
\vskip 2ex
\caption[Comparison of K-band Spectra]
{\label{osiris} OSIRIS K-band spectra of 2M2234+4041A (black) and B (gray) compared to a $\sim$1~Myr old Taurus M6 (CFHT-BD-Tau~18, blue) and a $\sim$1~Gyr field M6 (Gl406, magenta). Each spectrum is normalized by its median from 2.0 to 2.4~$\mu$m.  The  2.2~$\mu$m \ion{Na}{1} alkali lines in components A and B are much weaker than for the field dwarf, indicating that both components are young.  The continuum shape and 2.3 $\mu$m CO bandhead for components A and B are remarkably similar, in agreement with their identical spectral classification of M6.  
}
\end{figure}

We obtained spatially resolved $K$-band spectroscopy of \sciencebin\ on 2008 June 30 (UT) using the OSIRIS integral field spectrograph \citep{larkin03} and LGS~AO on the Keck-II telescope.  We selected the 20 mas/pixel scale for our observations.  
 We observed \sciencebin\ taking 8 dithered exposures of 120 seconds each, for a total of 16 minutes on source integration.   The FWHM, as measured on the stacked 2D images of \sciencebin, was 68$\pm$2~mas, thus the binary (158~mas separation) is well resolved. 
Immediately following our observations of \sciencebin, we obtained sky frames using the same dither pattern and integration time.  To correct for telluric absorption, we obtained spectra of a nearby A0V star, HD~209932.  The initial reduction from 2D images to 3D data cubes was accomplished using the OSIRIS data reduction pipeline \citep{krabbe04}.  The individual spectra for each component were then extracted from the 3D data cubes by summing the flux in fixed apertures of 175 $\times$ 175 mas at each wavelength.  The resulting spectra of each component were then median combined together.  Telluric correction and flux calibration were performed using the observations of the A0~V standard and the technique described in Vacca et al. (2003).  The resulting 1.96--2.38~$\mu$m spectra of \sciencebina\ and B (Figure \ref{osiris}) have a resolution ($\lambda/\Delta\lambda$) of $\sim$3500, and median S/N of $\sim$90 per pixel.

\subsection{Keck HIRES Optical Spectroscopy}
We acquired two optical spectra of \sciencebin\ on May 11 and August 12 of 2006 with the High Resolution Echelle Spectrometer (HIRES; Vogt et al.~1994) on the Keck I telescope.
We used the 0.861$\arcsec$ slit with HIRES to give a spectral resolution of $\lambda$/$\Delta\lambda$$\approx$58,000. 
To maximize the throughput near the peak of a M dwarf spectral energy distribution, we used the GG475 filter with the red cross-disperser. 

Each stellar exposure was bias-subtracted and flat-fielded for pixel-to-pixel sensitivity variations. After optimal extraction, the 1-D spectra were wavelength calibrated with a Th/Ar arc. Finally
the spectra were divided by a flat-field response and corrected to the heliocentric rest frame.
The final spectra were of moderate S/N reaching $\sim$ 25 per pixel at 7000 \AA.

\begin{figure}
\vskip 0.5in
\centerline{\includegraphics[width=6in,angle=90]{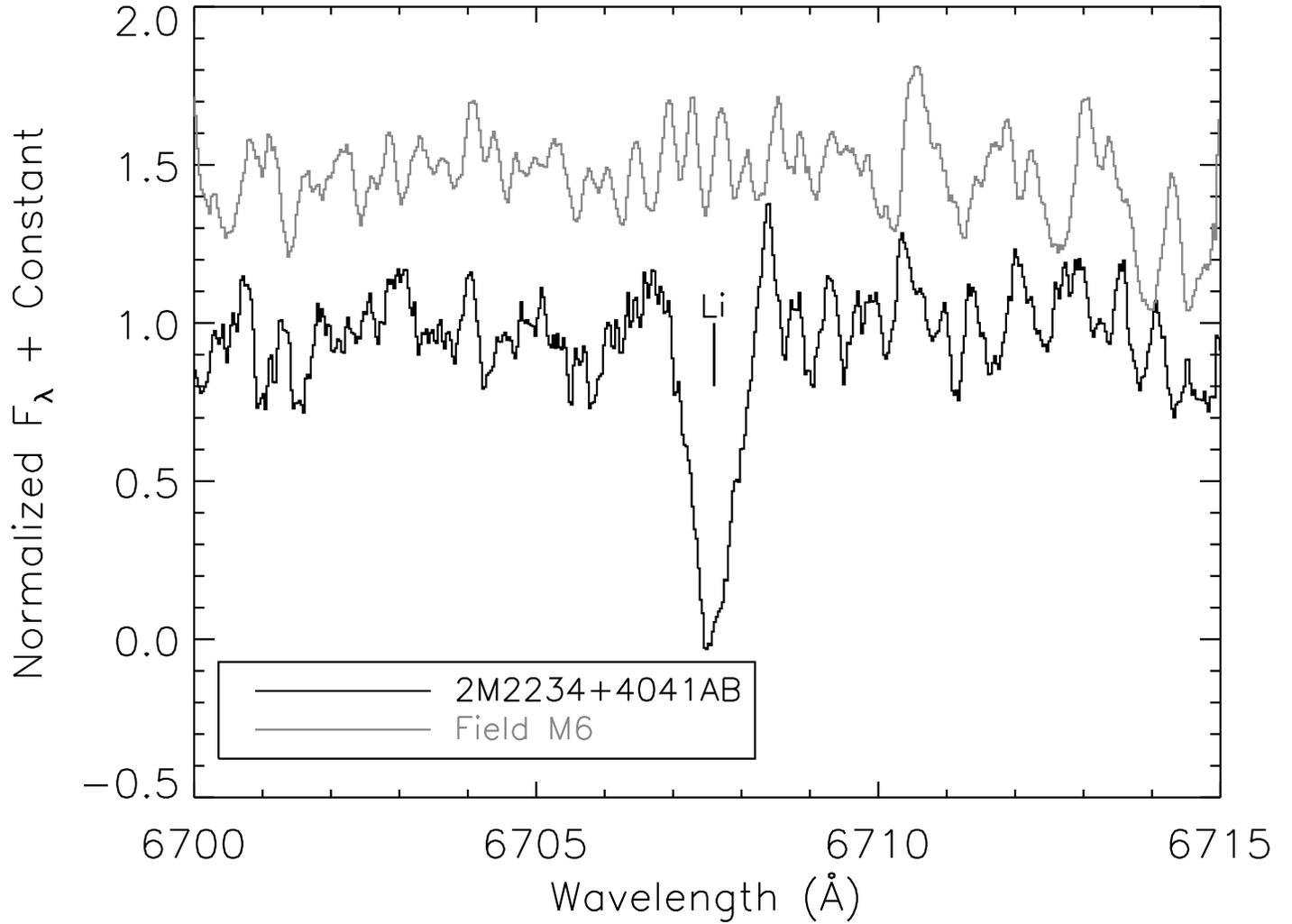}}
\vskip 2ex
\caption[Li]
{\label{li} HIRES spectrum of the lithium feature in \sciencebin\ compared with the M6 field dwarf NLTT~56194.  The detection of lithium in the spectrum of \sciencebin\ indicates that it is younger than the Pleiades ($\sim$100~Myr).}
\end{figure}

\begin{figure}
\vskip 0.5in
\centerline{\includegraphics[width=6in,angle=90]{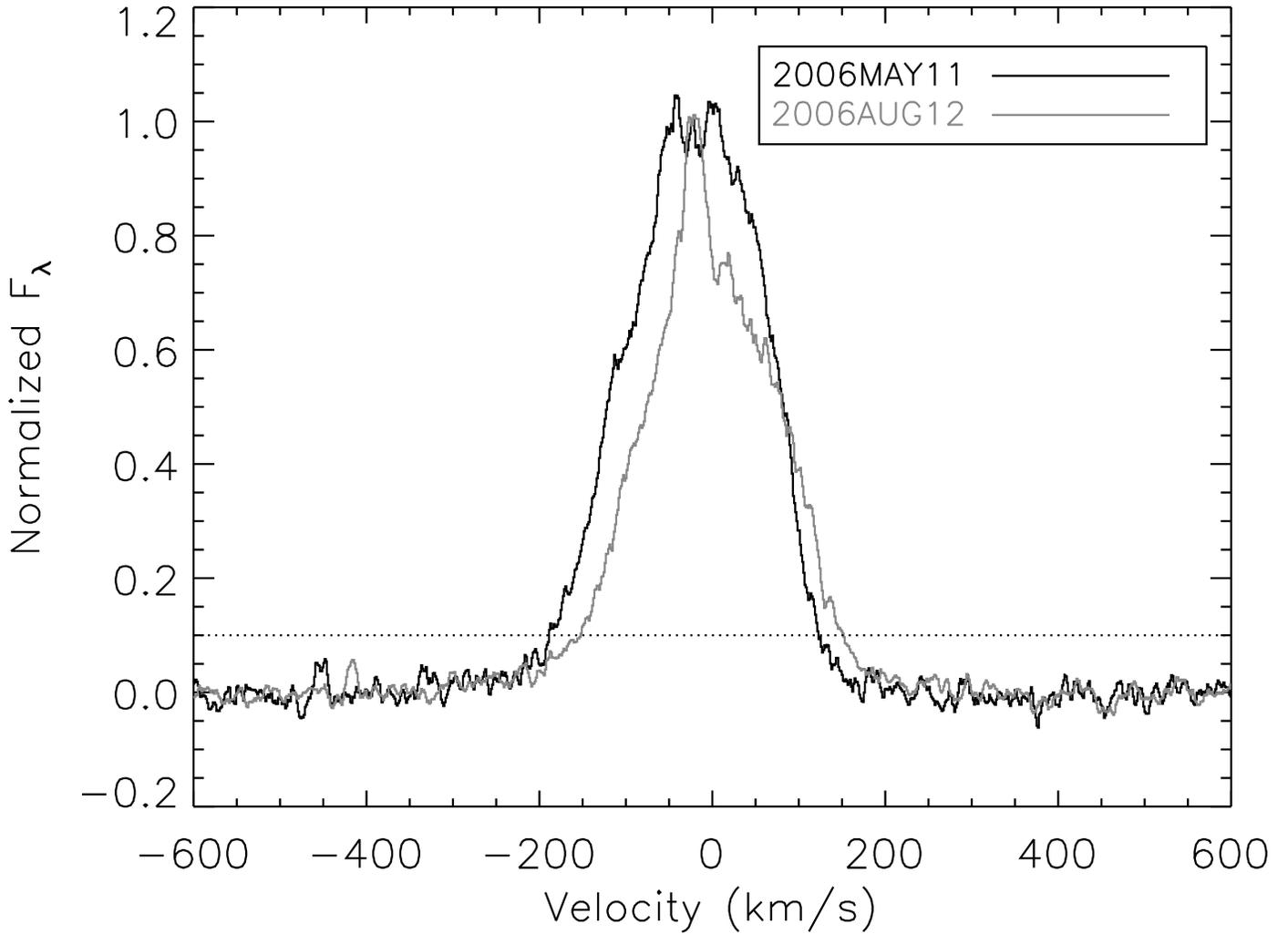}}
\vskip 2ex
\caption[Halpha]
{\label{halpha} H$\alpha$ velocity profile of \sciencebin\ from Keck/HIRES observations on 2006 May 11 (black) and 2006 Aug 12 (gray).  The profiles have been continuum-subtracted and normalized at the peak flux of the line.  The dotted line shows the 10\% flux level.  The H$\alpha$ line equivalents widths and 10\% widths imply ongoing mass accretion, and thus a young age for \sciencebin.}
\end{figure}

We cross-correlated each of 7 orders between 7000 and 9000 \AA\/ of each stellar spectrum with a RV standard of similar spectral type using IRAF's\footnote{IRAF (Image Reduction and Analysis Facility) is distributed by
the National Optical Astronomy Observatories, which is operated by the Association of Universities for Research in Astronomy, Inc.~(AURA) under cooperative agreement with the National Science Foundation.} {\it fxcor}
routine (Fitzpatrick 1993). We excluded the Ca II infrared triplet (IRT)\footnote{Since the target star exhibits Ca II emission that is not present in the RV standards, it was important to eliminate the IRT from the cross-correlation.} and regions of strong telluric absorption in the cross-correlation.  Radial velocites and relevant line equivalent widths measured from our HIRES data are listed in Table \ref{hires}, and portions of our HIRES spectrum are displayed in Figures \ref{li} and \ref{halpha}.  Further details of HIRES data reduction and analysis are provided by \citet{shkolnik09}.

\subsection{Proper Motion}
\begin{figure}
\epsscale{0.8}
\plotone{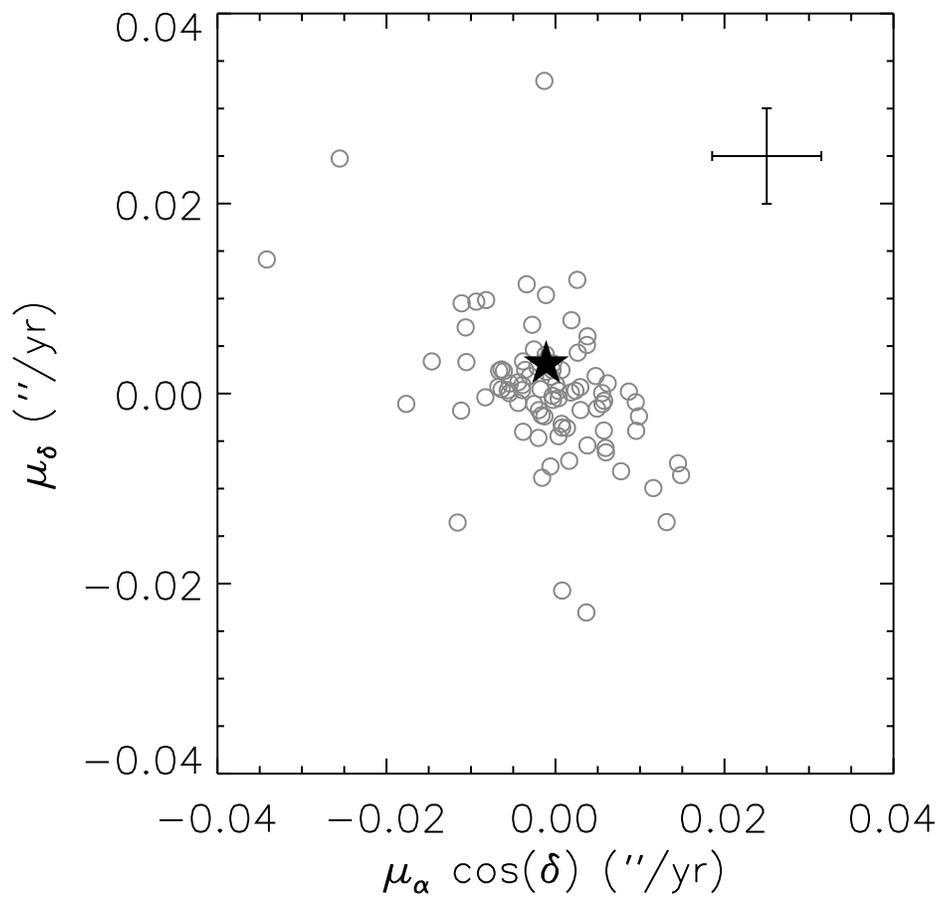}
\caption[Proper Motion]
{\label{pm} The proper motion of \sciencebin\ (filled star), as measured from POSS-I images (1952 July 20) and UH~2.2~m images (2007 July 31).  The gray open circles show the proper motions measured for objects matched in the two  epochs of images.  The uncertainty (shown in the upper right) is the standard deviation found when transforming the POSS-I detected positions to the Tek image coordinates, divided by the time baseline.}
\end{figure}

To calculate a proper motion for \sciencebin, we used red and blue POSS-I\footnote{The National Geographic Society - Palomar Observatory Sky Atlas (POSS-I) was made by the California Institute of Technology with grants from the National Geographic Society. }
images from the Digitized Sky Surveys\footnote{The Digitized Sky Surveys were produced at the Space Telescope Science Institute under U.S. Government grant NAG W-2166. The images of these surveys are based on photographic data obtained using the Oschin Schmidt Telescope on Palomar Mountain and the UK Schmidt Telescope. The plates were processed into the present compressed digital form with the permission of these institutions.} and 12 i' and z' band images from the Tektronix 2048$\times$2048 CCD camera (Tek) on the University of Hawaii 2.2-meter telescope kindly provided to us by Roy Gal.  \sciencebin\ is well detected ($>$30$\sigma$) in all of the POSS-I and Tek images.
The red and blue POSS-I images were taken on 1952 July 20 and have pixel scales of 1.7\arcsec\ and 1.0\arcsec\ per pixel respectively.  The Tek images were taken on 2007 July 31 and have a pixelscale of 0.22\arcsec\ per pixel.   We used Source Extractor \citep{bertin96} to obtain the x and y positions of objects in individual images and used IRAF's xyxymatch and geomap routines (with 3rd order polynomial fits) to match detections within each epoch and spatially transform the detections to the image coordinate system of the POSS-I red image (for POSS-I epoch images) or the first Tek image (for all Tek images).  The standard deviations of the fits found by geomap were $\sim$0.20\arcsec\ for the match of POSS-I red and blue images and $\sim$0.04\arcsec\ for the match of the Tek images.  We then calculated the average x and y positions within each epoch, and used xyxymatch and geomap to match 107 sources and transform their POSS-I detected positions to the Tek image (x,y) coordinate system, giving us the $\Delta$x and $\Delta$y of \sciencebin.  We then calculated the pixelscale and orientation of the Tek images using SCAMP (the Terapix astrometric software package), and converted $\Delta$x and $\Delta$y for \sciencebin\  to $\Delta\alpha$ and $\Delta\delta$.  The standard deviations found by geomap for transforming the POSS-I detections to the Tek image coordinate system were 0.34\arcsec\ in x and 0.27\arcsec\ in y.  The measured motion of \sciencebin\ from the POSS-I images to the Tek images is $-$58$\pm$340 mas and 166$\pm$270 mas respectively, resulting in a proper motion of $-$1$\pm$6 and 3$\pm$5~mas yr$^{-1}$ for $\mu_\alpha$(cos($\delta$)) and $\mu_\delta$ respectively. Thus, \sciencebin\ shows no significant proper motion (Figure \ref{pm}).

\section{Analysis}

\subsection{Spectral Type}
We derive an optical spectral type by visually comparing the TiO and VO bands in our HIRES spectrum of \sciencebin\ to standard stars of known spectral types.  For stars of spectral type $\sim$M5.5 or later, the 7140~\AA\ TiO band weakens due to condensation onto grains whereas the VO band at 7300~\AA\ strengthens with spectral type until M7, and then weakens for later types.   This allows us to unambiguously classify \sciencebin\ as M6 with a conservative uncertainty of $\pm$1 subclass.  The M6 optical spectral type we determine agrees with that of \citet{cruz03}, who used low-resolution (R$\simeq$1400) optical spectra.

We can also determine the spectral type of \sciencebin\ by comparing its SpeX near-IR spectrum to spectra of young objects in Taurus (Figure \ref{nirspt}).  We choose Taurus objects rather than field dwarfs for comparative spectral typing, as the spectral features of \sciencebin\ are consistent with a young age (see \S3.3).  Our near-IR spectrum of \sciencebin\ is a near perfect match to the spectrum of CFHT-BD-Tau~18, a young Taurus object with an optically determined spectral type of M6 \citep{guieu06}.  The spectral type sensitive index ($\langle F_{\lambda=1.550-1.560}\rangle/\langle F_{\lambda=1.492-1.502}\rangle$) of \citet{allers07}, which is calibrated for young and field stars with optical spectral types, suggests a spectral type of M6.4~$\pm$~1, in agreement with the spectral types we determine from visual inspection of optical and near-IR spectra.  We assign a spectral type of M6~$\pm$~1 to both components of \sciencebin\ given the similarity of their near-IR colors (Table~\ref{nirc2}) and K-band spectra (Figure~\ref{osiris}).

\subsection{Extinction}
To obtain intrinsic colors and magnitudes of \sciencebin\, we must determine the amount of dust attenuating the source.  To compute the extinction, we compare the observed J--H color of \sciencebin\ to the colors of field and young objects.  We choose to determine our extinction value using J--H for two reasons.  First, of the available photometry of \sciencebin\ (Table \ref{binphot}), the J-H color is at the shortest wavelength and hence is the most sensitive to extinction, and the least likely to be contaminated by excess emission from a circumstellar disk.  Second, the J--H color for mid to late M-type objects is insensitive to spectral type \citep{leggett02}, so uncertainties in the spectral type of \sciencebin\ do not affect our determination of extinction.  

The mean 2MASS J--H color of M6 V objects from Dwarfarchives\footnote{http://www.dwarfarchives.org} \citep{kirkpatrick91,kirkpatrick92,kirkpatrick95} is 0.58 mag with a standard deviation of 0.07 mag.  The 2MASS J--H color of \sciencebin\ is 0.74$\pm$0.03 mag.  Dereddening the J--H color of \sciencebin\ (0.74$\pm$0.03) to the field M6 V value, using (A$_J$ -- A$_H$)/A$_V$ = 0.1 (values from ADPS\footnote{The Asiago Database on Photometric Systems is available via http://ulisse.pd.astro.it/Astro/ADPS/}), requires A$_V$=1.6$\pm$0.8 mag.  This may represent an upper limit on A$_V$, however, as young objects may have redder intrinsic colors than older field objects \citep{kirkpatrick06}.  As a cross-check, we compiled 2MASS PSC photometry for 24 young M6 type objects in Upper Scorpius from \citet{ardila00} and \citet{slesnick06}.  At 5 Myr old, most of the gas and dust in the Upper Scorpius region has been dispersed by OB stars in the region.  Thus Upper Scorpius objects should have very low dust extinction, and provide a good estimate of the intrinsic colors of young objects.  The mean J--H color of the Upper Scorpius M6 objects is 0.69 mag, with a standard deviation of 0.07 mag, which implies A$_V$=0.5$\pm$0.8 for \sciencebin, which is in agreement (to within the uncertainties) with the A$_V$ we derive from comparison with field dwarf colors.  Based on the A$_V$'s and uncertainties from dereddening the J--H color of \sciencebin\ to field and young M6 objects, we adopt an extinction value of A$_V$=1.1$\pm$1.1 mag.

\subsection{Age}

\subsubsection{Lithium}
Lithium is destroyed at core temperatures of $2-3 \times 10^6$~K.  Such hot temperatures exist in the central regions of the lowest mass stars and high mass brown dwarfs \citep[$\gtrsim$ 0.06~$M_\odot$,][]{chabrier96}.  The timescale for lithium depletion is dependent on mass, with higher mass stars burning their lithium more quickly than lower mass stars.  
Figure \ref{li} shows lithium detected in the spectrum of \sciencebin, with an equivalent width of 0.83$\pm$0.02 \AA\ (the mean of the values listed in Table \ref{hires}).  \citet{stauffer98} found that Pleiades stars with spectral types earlier than M6.5 had depleted their lithium.  Thus, for a spectral type of M6, the detection of lithium in \sciencebin\ places an upper limit on the age of $\lesssim$100~Myr, based on the age of Pleiades \citep{meynet93}.

\subsubsection{\Lp -band Excess}
\begin{figure}
\vskip 0.5in
\centerline{\includegraphics[width=6in,angle=90]{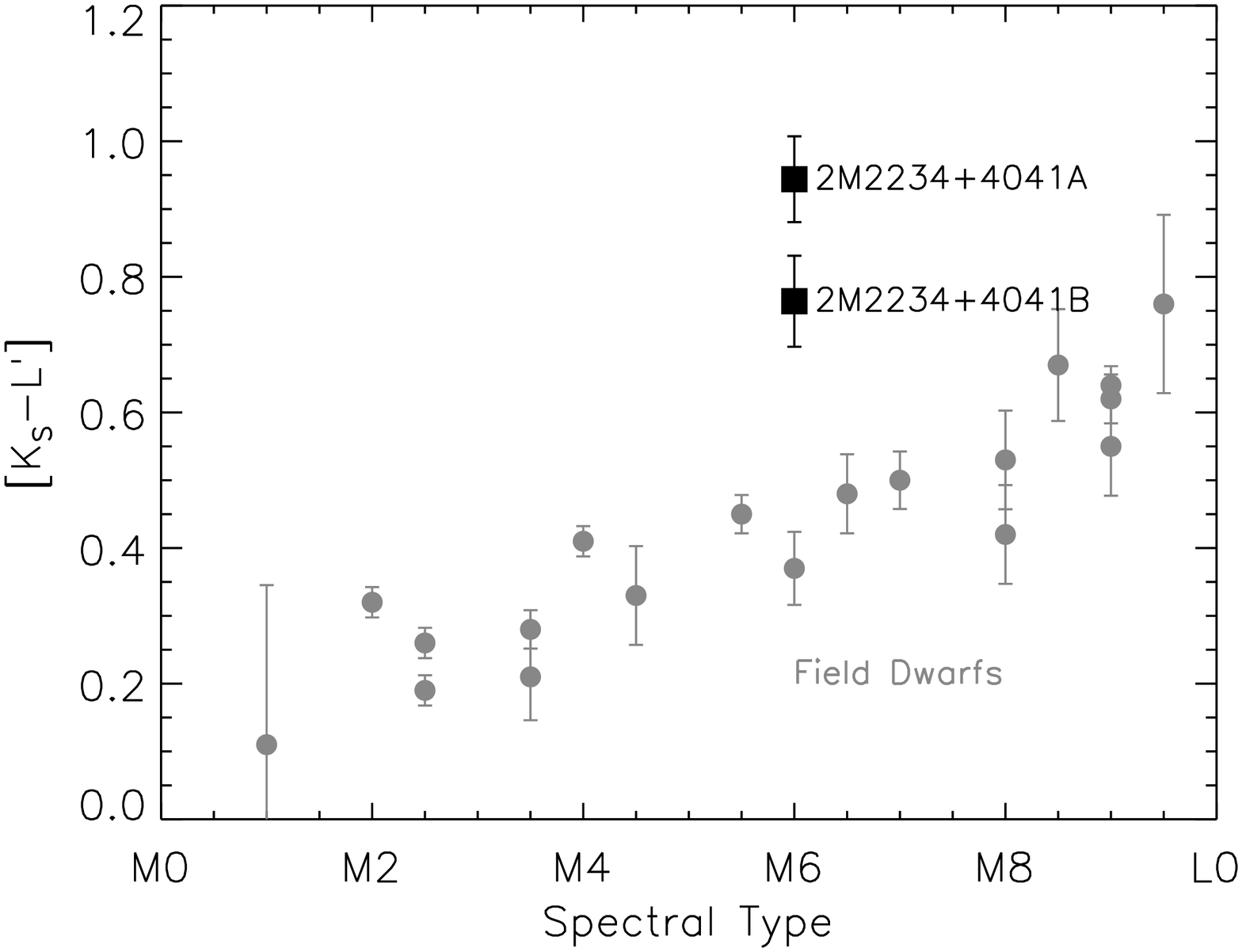}}
\vskip 2ex
\caption[Infrared Excess]
{\label{k-l} \Ks\ - \Lp\ colors of \sciencebina\ and B compared to field dwarfs.  The \Ks\ - \Lp\ colors of \sciencebina\ and B have been corrected for reddening (A$_V$=0.8) using (A$_{K_S}$ - A$_{L^{\prime}}$)/A$_V$ = 0.06.  The \Lp\ values for field dwarfs are from \citet{leggett02}, \citet{reid02}, and \citet{leggett03}, and \Ks\ is from 2MASS.  The redder \Ks\ - \Lp\ colors of \sciencebina\ and B, relative to field dwarfs of comparable spectral type, indicate that both components have infrared excesses, presumably due to the presence of circumstellar disks.  The uncertainty in \Ks\ - \Lp\ for \sciencebina\ and B is dominated by the \Lp\ calibration uncertainty. The relative difference in \Ks\ - \Lp\ color for the system, however, is known more precisely (0.18$\pm$0.04~mag), indicating different physical structures for the two components' disks.
}
\end{figure}

The detection of infrared emission in excess of that expected from the stellar photosphere is widely used as an indicator of circumstellar disks.  \citet{liu03} found that objects with spectral types as late as M8 can exhibit excess emission in \Lp\ band.  Figure \ref{k-l} shows the \Ks\ -- \Lp\ colors for \sciencebina\ and B compared to the colors of field dwarfs \citep{leggett02,reid02,leggett03}.  The dereddened \Ks\ -- \Lp\  colors of \sciencebina\ and B are 0.96$\pm$0.06 mag and 0.78$\pm$0.07 mag, whereas the average \Ks\ -- \Lp\ color of M5--M7 type field dwarfs is 0.45 mag.  Thus, \sciencebina\ and B have \Ks\ -- \Lp\  excesses detected at the 8 and 5 $\sigma$ levels, respectively, and likely possess circumstellar disks.  Half of primordial disks around stars dissipate by an age $\sim$3~Myr and all have dissipated by $\sim$6~Myr \citep{haisch01}.  There is some evidence that primordial disks around brown dwarfs might be slightly longer-lived than disks around stars \citep[e.g.][]{riaz08}, so it is difficult to place an upper limit on the age of \sciencebin\ based solely on the presence of its circumstellar disk.

It is important to note that while the absolute value of  \Ks\ -- \Lp\ for \sciencebina\ and B has an uncertainty of 0.06--0.07 mag, the uncertainty in their \emph{relative} \Ks\ -- \Lp\  colors is smaller (0.04 mag), so the difference in the \Ks\ -- \Lp\  colors of the two components is well detected ($\Delta$ \Ks\ -- \Lp\ =  0.18$\pm$0.04 mag).  The presence of circumstellar disks in both components of a wide, low-mass binary system is interesting, as such systems are rare \citep[e.g. ][]{allers07}.  Given that \sciencebina\ and B have nearly identical spectral types and $J,H,K_S$ brightnesses and are presumably co-eval, the difference in their excess is particularly compelling, as it implies that the physical structures of their disks are different.  Models of circumstellar disks \citep[e.g.][]{dalessio06,dullemond01} indicate that differences in the radius of an inner disk hole, the height of the disk rim, the accretion rate, or the inclination of disk could explain the relative \Ks\ -- \Lp\ colors of \sciencebina\ and B.  Unfortunately, with only \Ks\ -- \Lp\ colors, we can not distinguish between these scenarios.

\subsubsection{Accretion Indicators}
Young, low-mass objects often show spectroscopic signatures of accretion from circumstellar disks.  Atomic hydrogen lines, specifically H$\alpha$ emission, are commonly used to measure accretion rates.
Table \ref{hires} lists the H$\alpha$ equivalent widths (EWs) and 10\% widths measured in our HIRES spectra of \sciencebin.  \citet{white03} found that non-accreting stars and brown dwarfs can have strong H$\alpha$ emission (EWs of $-$3 to $-$36~\AA), presumably due to chromospheric activity, and proposed that M6--M7.5 type objects with H$\alpha$ EWs less than $-$40~\AA\ are likely accreting.  The H$\alpha$ EW we measure for \sciencebin\ in our 2006 May 11 spectrum is $-$52.7~\AA, well below the accretor limit, while the EW measured on 2006 Aug 12 ($-$39.2 \AA) is just above the limit.  Both of our H$\alpha$ EW measurements, however, meet the accretor criterion of \citet{barrado03}.  

\citet{white03} proposed a more reliable determination of accretion, the 10\% width of H$\alpha$ emission, which if greater than 270~km~s$^{-1}$ is indicative of accretion.  The H$\alpha$ 10\% widths we measure for \sciencebin\ (314 and 298~km~s$^{-1}$) are above the 10\% width requirement of \citet{white03} and well above the revised  200~km~s$^{-1}$ H$\alpha$ 10\% width cutoff for very low-mass stars and brown dwarfs adopted by \citet{mohanty05}.  Thus \sciencebin\ is actively accreting.  Following the procedure of \citet{natta04}, we calculate a mass accretion rate of 1.2 $\times 10^{-10} M_{\odot}~yr^{-1}$ from the average of our measured H$\alpha$ 10\% widths in Table~\ref{hires}.  The accretion rate we measure for \sciencebin\ is in good agreement with values for young M6 type objects in Taurus and Chamaeleon I determined by modeling the H$\alpha$ line emission profile \citep{muzerolle05}.
\citet{mohanty05} found that the fraction of accreting low-mass stars and brown dwarfs drops significantly for ages greater than 5~Myr.  Thus, the strong H$\alpha$ emission seen in the spectrum of \sciencebin\ implies that it is probably younger than 5~Myr.

The accretion rate we calculate from the H$\alpha$ equivalent width is measured from the {\it composite} HIRES spectra of \sciencebin.  
The OSIRIS K-band spectra of both \sciencebina\ and B show Br$\gamma$ emission (Figure~\ref{osiris}).  
\citet{muzerolle98} found that the luminosity of Br$\gamma$ line emission is well-correlated with the total accretion luminosity (determined from U-band or blue continuum emission) for late-K and early-M spectral type classical T~Tauri stars.  The relation between Br$\gamma$ and accretion luminosities was subsequently extended to higher and lower mass by \citet{calvet04} and \citet{natta06}.

To measure the Br$\gamma$ luminosities for \sciencebina\ and B, we flux calibrate our OSIRIS spectra using the Ks-band magnitude of each component, and then fit gaussians to the Br$\gamma$ line emission.  We measure Br$\gamma$ equivalent widths of $-$3.0$\pm$0.3~\AA\ and $-$2.5$\pm$0.3~\AA\ and calculate $log(L_{Br\gamma}/L_\odot)$ of $-$5.20$\pm$0.04 and $-$5.82$\pm$0.05 for \sciencebina\ and B respectively, assuming a distance of 325$^{+72}_{-50}$~pc to \sciencebin\ (see \S3.4).   
The uncertainties are determined from the formal errors of the gaussian fit and do not include distance uncertainties.  Using the relation between accretion luminosity and Br$\gamma$ luminosity from \citet{muzerolle98}, we calculate accretion luminosities, $log(L_{acc}/L_\odot)$, of $-$2.1$\pm$1.3 and $-$2.3$\pm$1.3 for \sciencebina\ and B.  Assuming a mass of 0.1~$M_\odot$ (\S3.5) and a radius of 0.982~$R_\odot$ (appropriate for a 1~Myr old, 0.1~$M_\odot$ object \citep{bcah98}), we calculate accretion rates (from $\dot{M}_{acc}=L_{acc} R_* / (GM_*)$) of 2.4 and 1.8~$\times~10^{-9}~M_\odot~yr^{-1}$ for \sciencebina\ and B respectively.  The accretion rates we measure from Br$\gamma$ are larger than the accretion rate we calculate from H$\alpha$, but given the large uncertainties in the Br$\gamma$ accretion luminosities and the distance to the source, this is not unexpected. The accretion rates we measure from Br$\gamma$ and H$\alpha$ are consistent with values of $\dot{M}$ for sources of similar mass \citep{muzerolle05}, and indicate that both components of \sciencebin\ are young and actively accreting.


\subsubsection{Gravity Sensitive Spectral Features}
\begin{figure}
\vskip 0.5in
\centerline{\includegraphics[width=6in,angle=90]{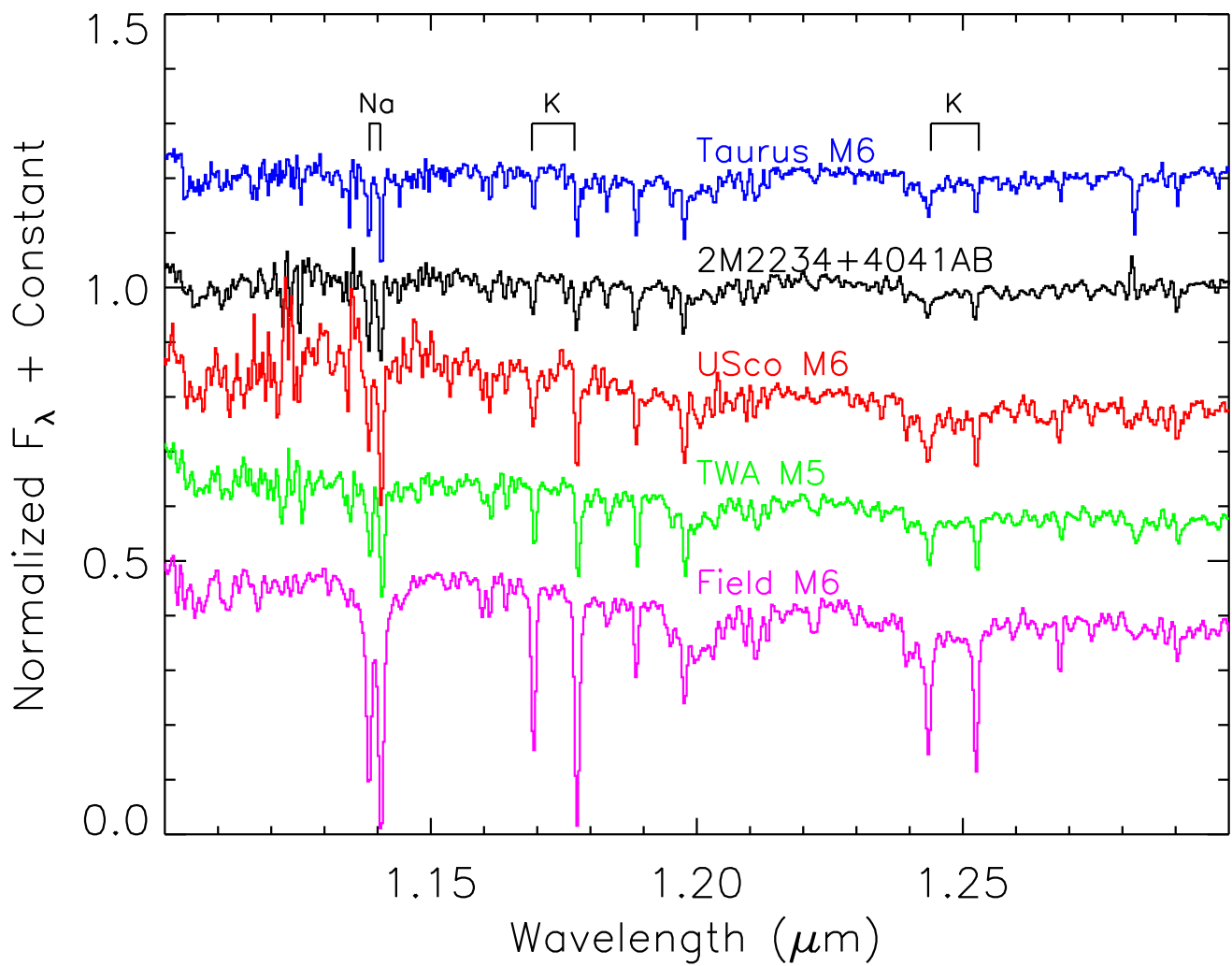}}
\vskip 2ex
\caption[Comparison of SpeX J-band Spectra]
{\label{spex} Integrated light J-band spectrum of \sciencebin\ (black) compared to a $\sim$1~Myr old Taurus M6 (CFHT-BD-Tau~18, blue), a $\sim$5~Myr old Upper Scorpius M6 (USco CTIO 75, red), a $\sim$10~Myr old TWHydra M5 (TWA8B, green), and a $\sim$1~Gyr field M6 (Gl406, magenta), all obtained with IRTF/SpeX.  Each spectrum is normalized by its median flux from 1.1--1.3~$\mu$m.  The depths of the \ion{K}{1} and \ion{Na}{1} alkali lines (labeled) increase with age. The depths of the alkali lines in the spectrum of \sciencebin\ agree quite well with the line depths seen in the Taurus M6, implying an age of $\sim$1~Myr for \sciencebin.}
\end{figure}

To further constrain the age of \sciencebin\, we examine the gravity (age) sensitive absorption lines of \ion{K}{1} and \ion{Na}{1} \citep{allers07,mcgovern04,gorlova03} in the SpeX J-band spectrum of \sciencebin.  For comparison to the \sciencebin\ spectrum, we selected objects with ages determined by their membership in: the Taurus star-forming region \citep[$\sim$1~Myr,][]{luhman03a}, the Upper Scorpius OB association \citep[$\sim$5~Myr,][]{preibisch02}, the TW Hydra moving group \citep[$\sim$10~Myr,][]{webb99}, and the general field dwarf population ($\gtrsim$1~Gyr).   We chose objects with optical spectral types of M6 (if available), so that we are comparing only the age sensitivity of spectral features.  In Figure \ref{spex}, we compare the SpeX spectrum of \sciencebin\ to SpeX spectra we obtained of CFHT-BD-Tau~18 \citep[Taurus M6,][]{guieu06}, USco~CTIO~75 \citep[Upper Scorpius M6,][]{ardila00}, as well as a SpeX spectrum of TWA8B \citep[TW Hydra M5,][]{webb99} given to us by Casey Deen ({\it priv.~comm.}), and a spectrum of Gl406, a field M6 from \citet{cushing05}.  In Figure \ref{spex}, one can clearly see the increase in \ion{K}{1} and \ion{Na}{1} line strengths with age.  The line strengths in \sciencebin\ are noticeably weaker than the field, TW Hydra, and Upper Scorpius objects, indicating an age for \sciencebin\ of less than 5~Myr.  The  \ion{K}{1} and \ion{Na}{1} line strengths (and the entire spectrum) of \sciencebin\ are best matched by CFHT-BD-Tau~18, implying an age of $\sim$1~Myr for \sciencebin.  The \ion{K}{1} 7700\AA\ line in the composite HIRES spectrum is also very weak and indicative of a young age \citep{shkolnik09}. Our resolved OSIRIS K-band spectra of \sciencebina\ and B have 2.2 $\mu$m \ion{Na}{1} line strengths similar to a Taurus M6 (Figure \ref{osiris}), further indicating that both components of \sciencebin\ are young.

\subsection{Distance}
The distance to \sciencebin\ is the critical parameter for determining the physical separation of the system, as well as its luminosity and mass.
Assuming \sciencebin\ is of similar age as objects in Taurus (\S3.3), we can use the absolute magnitudes of Taurus objects to estimate a photometric distance for \sciencebin.  The absolute J-band magnitudes of M5.5-M6.5 type objects in Taurus \citep{briceno02,luhman03a,luhman04b,guieu06} range from 4.57 to 7.39 mags with a mean of 6.51 mags.  The absolute J-band magnitudes include correction for extinction, exclude objects listed as binaries in the references above or discovered as binaries in \citet{konopacky07} and \citet{kraus06}, and assume a distance to Taurus of 143~pc \citep{loinard08}.   The dereddened J-band magnitude of \sciencebina\ is 13.00~$\pm$0.30 mag (accounting for extinction uncertainties), implying a distance modulus of 6.49 and yielding a photometric distance of 199~pc.  The range of absolute J-band magnitudes for Taurus objects corresponds to distance moduli of 5.61--8.43 mags, so distances of 132--484~pc for \sciencebin\ are plausible.

\begin{figure}
\vskip 0.5in
\centerline{\includegraphics[width=6in]{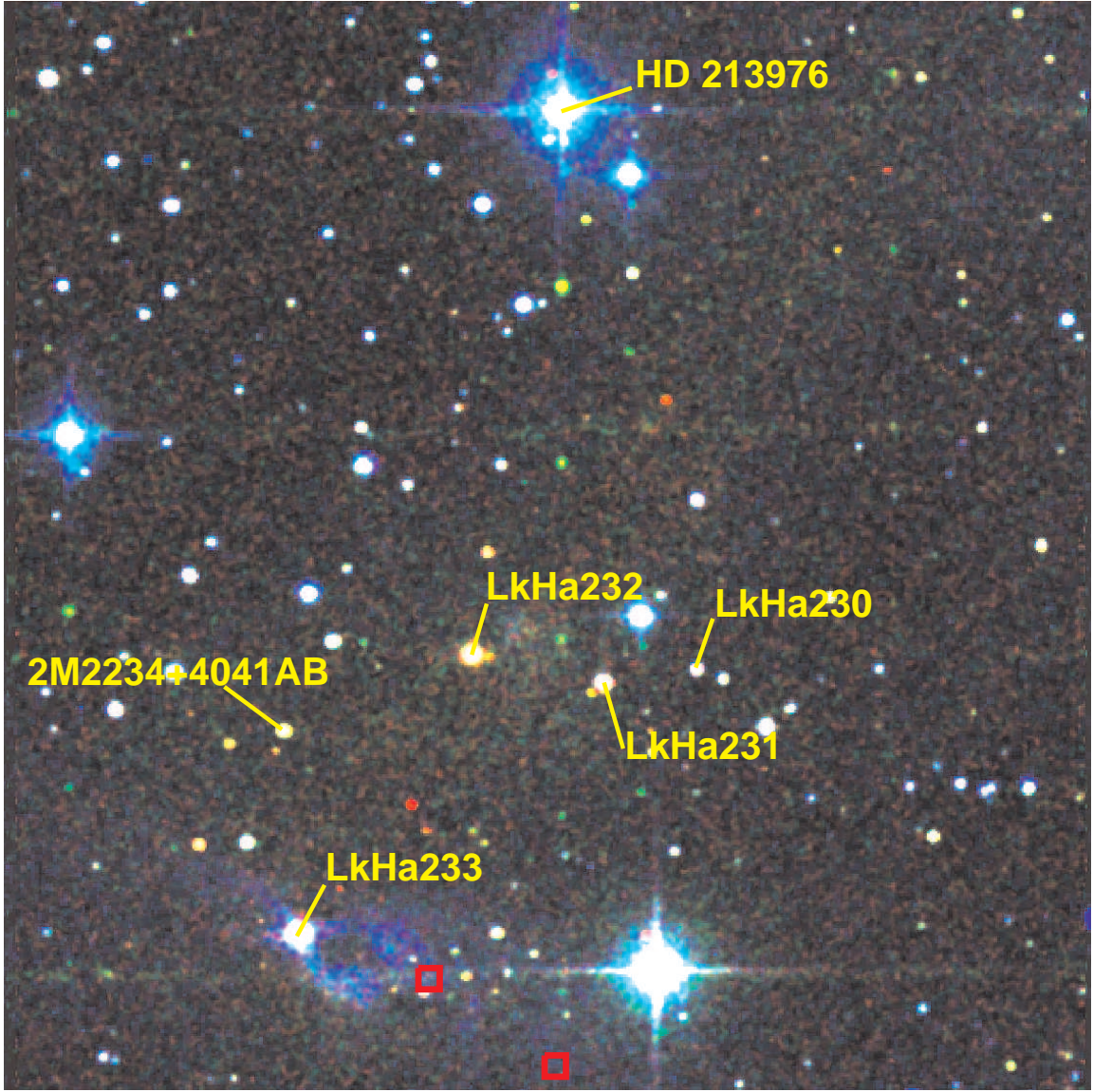}}
\vskip 2ex
\caption[Image of region]
{\label{region} Three-color image showing the location of \sciencebin\ relative to other young stars in the region.  The color mapping is red for 2MASS \Ks -band, green for 2MASS $J$-band, and blue for POSS-I-Red images.  Herbig-Haro objects from \citet{mcgroarty04} are displayed as red squares.}
\end{figure}

Given its young age ($\sim$1~Myr), null proper motion measurement and low radial velocity (Table \ref{binprops}), where could \sciencebin\ have originated?
\sciencebin\ lies 1.5\arcmin\ from the well studied Herbig Ae/Be star, LkH$\alpha$233 \citep{calvet78}, and near the young K-type stars, LkH$\alpha$ 230, 231 and 232, and the B star, HD~213976 (Figure \ref{region}).  Based on the geometry and temperature derived from observations of molecular gas (CO, NH$_3$, and H$_2$CO) in the region, \citet{olano94} concluded that the LkH$\alpha$ stars likely formed from Condensation A of the LBN~437 complex, which is externally heated by HD~213976.  HD~213976 has a \emph{Hipparcos} measured parallax of 3.08$\pm$0.56 mas \citep{vanleeuwen07}, corresponding to a distance of 325$^{+72}_{-50}$~pc.  If at this distance, the projected separation of \sciencebin\ is 51$^{+11}_{-8}$~AU.

One method of establishing membership with a group of stars is common space motion.  The null proper motion we measure for \sciencebin\ is consistent with the \emph{Hipparcos} measured proper motion of HD~213976 \citep[$-$1.67$\pm$0.46, $-$3.06$\pm$0.45 mas/yr,][]{vanleeuwen07}.  The proper motions reported by \citet{ducourant05} for LkH$\alpha$230 (4$\pm$7, $-$11$\pm$3 mas/yr), LkH$\alpha$231 ($-$9$\pm$5, $-$8$\pm$5 mas/yr), and LkH$\alpha$232 ($-$11$\pm$5, $-$3$\pm$5 mas/yr) are also low.  Our measured radial velocity of \sciencebin\ is $-$10.6~$\pm$~0.5~km~s$^{-1}$ (Table \ref{binprops}), which agrees with the radial velocity of LkH$\alpha$233 inferred from the relative velocities of its red and blue shifted jets \citep[$\sim$$-$10~km~s$^{-1}$,][]{perrin07}.
The radial velocity of HD~213976 ($-$17.2~km~s$^{-1}$) agrees with that of \sciencebin\ to within the $\sim$10~km~s$^{-1}$ spread observed for stars in the Taurus star-forming region \citep{bertout06}.  Kinematically, \sciencebin\ could be associated with the LkH$\alpha$233 group.

The distance typically assumed for the LkH$\alpha$233 group of young stars is 880~pc, a photometric distance of HD~213976 dating back to \citet{herbig60}.  Photometric distances of 700 and 600~pc were found for LkH$\alpha$231 and 232 \citep{calvet78}, and \citet{chernyshev88} determined a photometric distance of 660~pc for 21 emission stars in the region.  Given the large discrepancy between the astrometric distance (325$^{+72}_{-50}$~pc) and the original photometric distance (880~pc) of HD~213976, it is worthwhile to re-examine the photometric distance to the LkH$\alpha$233 group and see if the discrepancy can be resolved.


Calculating a photometric distance for LkH$\alpha$233 is difficult due to the fact that its circumstellar disk is viewed edge-on, and thus LkH$\alpha$233 is likely seen in scattered light \citep{perrin07}.  We can, however, obtain photometric distances for HD~213976, LkH$\alpha$232 and LkH$\alpha$231. 
\citet{hernandez05} report an optical spectral type of B3 for HD~213976 and a low extinction (A$_V=0.24 \pm 0.12$ mag).  To obtain a photometric distance to HD~213976, we compare its brightness to B2--B4 type stars in Upper Scorpius.  The mean absolute J-band brightnesses of twelve B2--B4 type Upper Scorpius members \citep{hernandez05} is $-$0.72$\pm$0.73 mag \citep[for $d=$145~pc;][]{dezeeuw99}.  By comparison to the brightness of Upper Scorpius objects, the 2MASS J-band magnitude of HD~213976 (7.19$\pm$0.02 mag) implies a distance modulus of 7.91 mag and a distance of 382$^{+153}_{-109}$~pc, in good agreement with the \emph{Hipparcos} distance.  Similarly, one can compare the brightnesses of LkH$\alpha$232 and LkH$\alpha$231 \citep[SpTs of K3 and K4 respectively,][]{herbig88} to young Taurus objects of similar spectral type from \citet{kenyon95}.  The mean absolute 2MASS J-band flux of K2--K5 type stars in Taurus (corrected for reddening using the published A$_V$'s of \citet{kenyon95}, and assuming a distance of 143~pc \citep{loinard08}) is 3.80$\pm$1.17 mag.  Objects similar to Taurus K2--K5 stars at 325~pc should then have observed J-band magnitudes of 11.35$\pm$1.17, in good agreement with the 2MASS J-band photometry of LkH$\alpha$232 and 231 (11.18$\pm$0.03 and 11.74$\pm$0.02 mag respectively).  
Our newly calculated photometric distances of HD~213976, LkH$\alpha$232 and LkH$\alpha$231 are in good agreement with the astrometric distance of 325$^{+72}_{-50}$~pc, which we adopt for the LkH$\alpha$233 group and \sciencebin.  This distance is similar to the distance of Lacerta~OB1 \citep[368$\pm$17~pc][]{dezeeuw99}, which lies $\sim$5 degrees ($\sim$55~pc) away, and may indicate that the LkH$\alpha$233 group is a site of recent star-formation associated with Lac OB1.   

\subsection{Luminosity}
To calculate the bolometric luminosities of \sciencebina\ and B, we apply a bolometric correction ($BC$) to their extinction-corrected near-IR magnitudes.
In theory, bolometric corrections can be made at any wavelength, but because \sciencebina\ and B harbor circumstellar disks (\S3.3.2 and 3.3.3), we choose to apply a bolometric correction at J-band ($BC_J$) to minimize possible contamination by circumstellar disk emission.  To find an appropriate $BC_J$, we tabulated 2MASS J-band magnitudes and $m_{bol}$ of field M5.5--M6.5 type dwarfs from \citet{leggett00}.  The mean $BC_J$ is 2.00 mag with a standard deviation of 0.06 mag.  For a distance of 325$^{+72}_{-50}$~pc, and a solar M$_{bol}$ of 4.76 mag, we calculate luminosities ($log(L_\star / L_\odot)$) of  $-$0.27$\pm$0.21 and $-$0.30$\pm$0.21 dex for \sciencebina\ and B, where the uncertainties were calculated from the uncertainties in A$_V$, distance, photometry, and $BC_J$.


\subsection{Mass}
\begin{figure}
\vskip 0.0in
\centerline{\includegraphics[width=4in,angle=0]{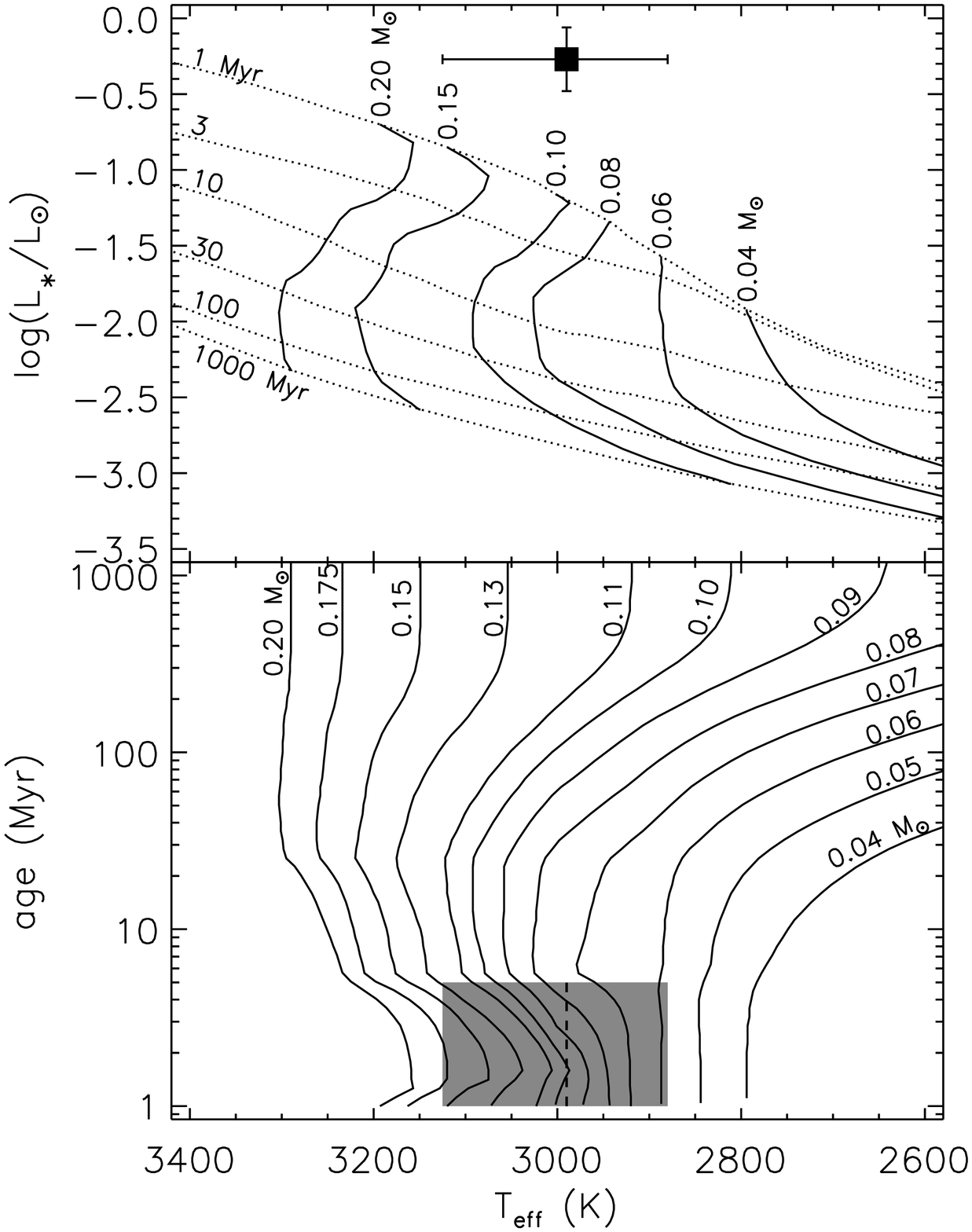}}
\vskip 2ex
\caption[Mass Determination]
{\label{mass} Mass determination for \sciencebin.  Top:  HR-diagram for \sciencebina.  Evolutionary models of \citet{bcah98} are overlaid, with isochrones plotted as dotted lines and iso-mass contours plotted as solid lines.  The square shows the position of \sciencebina\ on this diagram, with the luminosity calculated in \S3.5.  The T$_{eff}$ of \sciencebin\ is determined from its spectral type (M6$\pm$1) using the SpT-T$_{eff}$ relation of \citet{luhman03b}.  \sciencebina\ and B, having identical SpTs (and hence T$_{eff}$'s) and similar luminosities, are indistinguishable on this diagram.  \sciencebina\ is more luminous than the youngest isochrone in this diagram, making a mass estimate difficult. 
Bottom: Effective temperature vs. age diagram.  Iso-mass contours are from the evolutionary models of \citet{bcah98}.  The T$_{eff}$ of \sciencebin\ is displayed as a dashed line, and the shaded region shows the likely age and T$_{eff}$ range of \sciencebin.  The age range ($<$5~Myr) is based on the presence of a circumstellar disk, evidence of accretion, and the strength of the \ion{K}{1} and \ion{Na}{1} lines (see \S3.3 for details).  Fortunately, the iso-mass contours at young ages are nearly vertical (age insensitive).  Thus, we determine a mass for \sciencebina\ and B of 0.100~M$_\odot$, but masses of 0.060--0.175~M$_\odot$, are possible.
}
\end{figure}

Masses (and ages) of low-mass stars and brown dwarfs can be determined from evolutionary models by placing them on an H-R diagram with the objects' luminosities determined photometrically and effective temperatures determined spectroscopically.  The spectral type of \sciencebin, M6$\pm$1, corresponds to an effective temperature of 2990$^{+135}_{-110}$~K according to the SpT-T$_{eff}$ relationship appropriate for young objects of \citet{luhman03b}.  The top panel of Figure \ref{mass} shows the position of \sciencebina\ on an H-R diagram with the evolutionary models of \citet{bcah98} overlaid.  \sciencebina\ lies well above the 1~Myr old isochrone, making it difficult to estimate its mass using this method, though it is not uncommon for young objects to be overluminous relative to evolutionary models \citep[e.g. ][]{luhman04b}.

Given the well established age of $<$5~Myr, we can estimate the mass of \sciencebin\ from evolutionary models using {\it age} and effective temperature.  The bottom panel of Figure \ref{mass} shows the effective temperature and age range of \sciencebin, with iso-mass contours from \citet{bcah98} overlaid.  Fortunately, within the effective temperature range of \sciencebin, the iso-mass contours are nearly vertical, or age independent.  At an age of 1--2~Myr, the T$_{eff}$ of \sciencebin\ is closest to the 0.100~M$_\odot$ iso-mass track, but masses of 0.060--0.175~M$_\odot$ are also within the range of uncertainty in effective temperature and age.  Though the possible mass of \sciencebin\ falls in a wide range, it is the first potential VLM member of the LkH$\alpha$233 group.

\subsection{Orbital Period}
We have measured the projected separation of \sciencebin\ to be
51$^{+11}_{-8}$~AU, but the semimajor axis of the binary depends on its
orbital parameters.  Following the method of
\citet{torres99}, we assume random viewing angles and a
uniform eccentricity distribution between 0~$<~e~<$~1 to derive a
correction factor of 1.10$^{+0.91}_{-0.36}$ (68.3\% confidence limits)
for converting the projected separation into a semimajor axis.  This results in a semimajor axis of 57$^{+51}_{-21}$~AU.  For a total mass of 0.2~M$_\odot$, this corresponds
to an orbital period of 1000$^{+1600}_{-500}$~years.  Unfortunately, the orbital period of \sciencebin\ is far too long for our observations to detect any significant relative motion of the binary.


%
%
\subsection{Could \sciencebin\ be a Higher Order Multiple System?}
\begin{figure}
\vskip 0.5in
\centerline{\includegraphics[width=6in,angle=90]{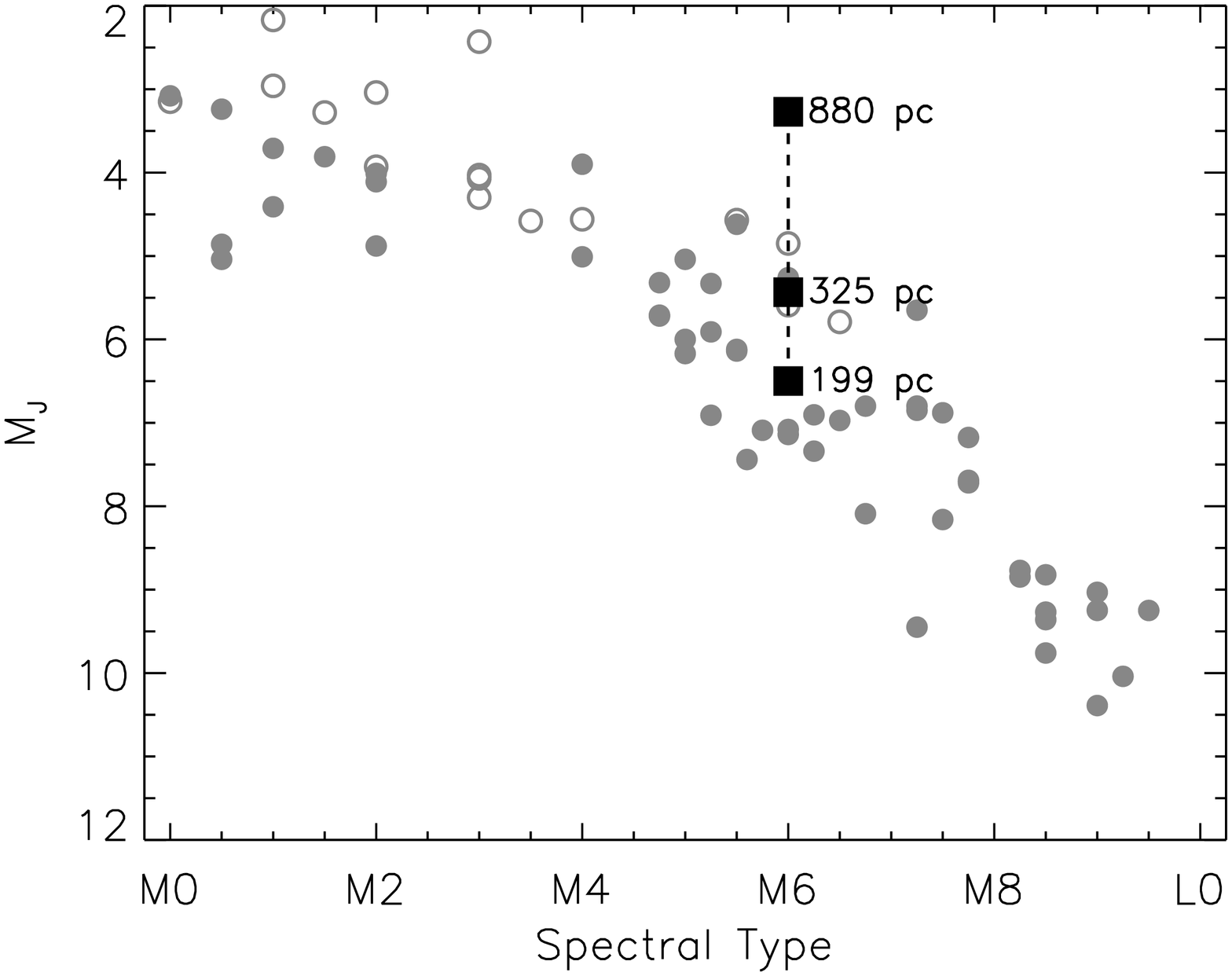}}
\vskip 2ex
\caption[Distance Determination]
{\label{hr} Spectral type vs. absolute J-band magnitude for \sciencebina\ (black squares) and Taurus objects \citep[gray circles,][]{briceno02, luhman03a,guieu06}.  The open gray circles are objects listed as binaries in the references above, \citet{kraus06}, or \citet{konopacky07}.  The black squares show the absolute J-band magnitude of \sciencebina\ if it resides at: the distance typically assumed for LkH$\alpha$233 group \citep[880~pc;][]{calvet78}, the Hipparcos distance to the nearby B3 star, HD213976  \citep[325~pc;][]{vanleeuwen07}, and the photometric distance to \sciencebina\ (calculated from the mean absolute J-band magnitude of M6 type Taurus objects).  See \S 3.4 for details.}
\end{figure}

As noted in \S3.4 and shown in Figures \ref{hr} and \ref{mass}, \sciencebina\ is over-luminous relative to young Taurus M6 objects and the theoretical 1~Myr old isochrone.  A common explanation for over-luminosity is unresolved multiplicity.  Is it possible that \sciencebin\ is a higher order multiple system?  Both components would need to have similar companions (i.e. a quadruple system) since the magnitudes and colors of \sciencebina\ and B are nearly identical.  We see no evidence of additional companions in three epochs of imaging (Table \ref{nirc2}).  Our LGS~AO observations, however, would only be able to detect companions with separations $\gtrsim$13~AU, whereas most field VLM binaries have tighter separations \citep{burgasser07}.  Neither epoch of HIRES data show evidence of spectroscopic binarity, and the radial velocity shows no sign of variation.  Interferometric or higher spatial resolution observations of the source would be needed to definitively rule out a higher order multiple system.

\subsection{Will \sciencebin\ be Disrupted?}
The 51~AU projected separation of \sciencebin\ is significantly wider than those typical of field VLM binaries \citep[3--15~AU;][]{burgasser07}, which may indicate that binary systems similar to \sciencebin\ are disrupted before they reach the age of the field population (a few Gyr).  In recent years, a handful of young, very wide ($>$100~AU) binaries have been discovered \citep{bejar08, bouy06, close07, luhman04a}.  \citet{close07}  suggest that such young, wide systems might become unbound due to dynamical interactions with other young stars in their natal clusters.  The separation and total mass of \sciencebin, however, meet the stability criteria of \citet{close07}, meaning that systems similar to \sciencebin\ are likely to survive the dispersion of their cluster and are unlikely to be disrupted by encounters with field stars.  Even if the actual masses of \sciencebina\ and B were at the low-end of their possible mass range (0.06~M$_\odot$), the system would still be considered stable.  The binding energy (GM$_1$M$_2$/a) of \sciencebin\ is $\sim$ 3.4~$\times 10^{42}$~erg which is lower than all but two (out of 69) known field VLM binaries, indicating that systems similar to \sciencebin\ are rare, but possible among the field population.

\section{Conclusions}
As a part of our Keck LGS~AO survey of young field objects, we discovered \sciencebin\ to be a binary.  Over three epochs of imaging, the separation of the system does not change, indicating that \sciencebina\ and B are physically associated (though the low proper motion of the system makes it difficult to confirm if the pair is actually co-moving).  
The similarity in spectral shape and features as well as the fluxes and colors of \sciencebina\ and B, suggest that both components have the same mass, spectral type, and age.
From composite optical and near-IR spectroscopy, we derive a spectral type of M6 for the system.  
Evidence of the youth of the system is the detection of \Ks\ -- \Lp\ excesses for \sciencebina\ and B (presumably due to the presence of circumstellar disks).  The level of excess, however, differs for the two components, indicating that the physical structures of their disks must be different.  The strengths of H$\alpha$ and Br$\gamma$ emission lines signify ongoing accretion, in agreement with the young ($\sim$1~Myr) age we derive for the system based on comparison of near-IR gravity sensitive spectral features with objects of similar spectral type and known age.  By constraining the age and T$_{eff}$ (from its spectral type), we determine individual masses of 0.10$^{+0.075}_{-0.04}$ M$_\odot$ for \sciencebina\ and B by comparison to evolutionary models \citep{bcah98}.

The original identification of \sciencebin\ was as a young, field object, not associated with any known star-forming regions \citep{cruz03}.  However, \sciencebin\ lies 1.5\arcmin\ away from the well-studied HAeBe star, LkH$\alpha$233.  LkH$\alpha$233 belongs to a small group of young stars (including LkH$\alpha$232, 231 and HD~213976), historically thought to lie at a distance of 880~pc \citep[the photometric distance to HD213976;][]{herbig60}.  Recently, \citet{vanleeuwen07} reported a \emph{Hipparcos} parallax measurement of 3.08$\pm$0.56 mas for HD~213976, corresponding to a distance of 325$^{+72}_{-50}$~pc.  We re-examined the photometric distance to HD213976 as well as LkH$\alpha$231 and 232.  Using the absolute J-band magnitudes of young objects (Taurus or Upper Scorpius), we derive photometric distances that are in agreement with the astrometric distance of 325~pc.  
\sciencebin\ is the first potential VLM member of the group.
At a distance of 325~pc, the physical separation of \sciencebin\ is 51~AU.  It is interesting that the first VLM object discovered in the region happens to be a young, wide binary.  The binding energy of \sciencebin, indicates that it is unlikely to be dynamically disrupted \citep{close07}.  

\acknowledgments
We gratefully acknowledge the Keck LGS AO team for their exceptional
efforts in bringing the LGS AO system to fruition.  It is a pleasure
to thank Randy Campbell, Jim Lyke, Al Conrad, Hien Tran, Christine Melcher, Cindy Wilburn, Joel Aycock, Jason McElroy and the Keck Observatory staff for assistance
with the observations.  This work is based in part on data products produced at the TERAPIX data center located at the Institut d'Astrophysique de Paris.  We also thank John Rayner and the IRTF Observatory staff for assistance with our SpeX observations and data reduction.
We are appreciative of conversations with George Herbig on the origin of \sciencebin\ and the nature of other nearby young stars.  We also thank Jeffrey Rich for translating \citet{chernyshev88} into English.
The authors sincerely thank Roy Gal for providing Tek/UH88" imaging of \sciencebin.  
We are grateful to Dan Jaffe, Casey Deen, Jasmina Marsh and Alan Tokunaga for providing the near-infrared spectrum of TWA8B.  This research has benefitted from the M, L, and T dwarf compendium housed at DwarfArchives.org and maintained by Chris Gelino, Davy Kirkpatrick, and Adam Burgasser.  KLC is supported by NASA through the Spitzer Space Telescope Fellowship Program, through a contract issued by the Jet Propulsion Laboratory, California Institute of Technology under a contract with National Aeronautics and Space Administration.  MCL, KNA and TJD acknowledge support for this work from NSF grant AST-0507833 and AST-0407441.  MCL also acknowledges support from an Alfred P. Sloan Research Fellowship.  ES ackowledges support from NASA/GALEX grant NNX07AJ43G.  KNA was partially supported by NASA Origins of Solar Systems grant NNX07AI83G.

\clearpage

\begin{deluxetable}{lccccccc}
\tablecaption{Keck LGS AO Observations}
\tabletypesize{\small}
\tablewidth{0pt}
\tablehead{
  \colhead{Date} &
  \colhead{Filter\tablenotemark{a}} &
  \colhead{Airmass} &
  \colhead{FWHM} &
  \colhead{Strehl ratio} &
  \colhead{Separation} &
  \colhead{Position angle} &
  \colhead{$\Delta$mag} \\
  \colhead{(UT)} &
  \colhead{} &
  \colhead{} &
  \colhead{(mas)} &
  \colhead{} &
  \colhead{(mas)} &
  \colhead{(deg)} &
  \colhead{}
}
\startdata

2006-Oct-14 & \Ks\ & 1.50 & 78 $\pm$ 5     & 0.18 $\pm$ 0.02    & 158.5 $\pm$ 0.6 & 99.3 $\pm$ 0.3    & 0.05 $\pm$ 0.03  \\

2007-Sep-06 & $J$  & 1.08 & 37.2 $\pm$ 0.8 & 0.116 $\pm$ 0.005  & 158.2 $\pm$ 0.3 & 99.52 $\pm$ 0.09  & 0.06 $\pm$ 0.03  \\
            & $H$  & 1.07 & 40.8 $\pm$ 0.7 & 0.161 $\pm$ 0.008  & 158.5 $\pm$ 0.3 & 99.53 $\pm$ 0.15  & 0.096 $\pm$ 0.010  \\
   
2008-Sep-08 & \Lp\ & 1.07 & 93 $\pm$ 3     & 0.58 $\pm$ 0.07    & 158.2 $\pm$ 1.0 & 100.1 $\pm$ 0.3  & 0.23 $\pm$ 0.03  \\

\enddata

\tablenotetext{a}{All photometry on the MKO system.}

\tablecomments{The tabulated uncertainties are the RMS of the
  measurements.  The astrometric errors are computed by appropriately
  combining in quadrature: (1) the instrumental measurements from
  fitting the images of the binary (with errors derived from fitting
  of simulated binary images) and (2) the overall uncertainties in the
  NIRC2 pixel scale and orientation.  See \S~2.1 for details.}
\label{nirc2}
\end{deluxetable}

\begin{deluxetable}{llll}
\tablecolumns{4}
\footnotesize
\tablecaption{Photometry of \sciencebin}
\tablewidth{0pt}
\tablehead{
\colhead{Band}                  &
\colhead{A+B}             &
\colhead{A\tablenotemark{a}}             &
\colhead{B\tablenotemark{a}}             \\
\colhead{}                  &
\colhead{(mag)}             &
\colhead{(mag)}             &
\colhead{(mag)}             
}
\startdata 
$J$ & 12.57$\pm$0.02\tablenotemark{b}   & 13.29$\pm$0.02 & 13.35$\pm$0.03 \\
$H$ & 11.83$\pm$0.02\tablenotemark{b}   & 12.54$\pm$0.02 & 12.63$\pm$0.02 \\ 
\Ks\ & 11.44$\pm$0.02\tablenotemark{b}  & 12.17$\pm$0.02 & 12.22$\pm$0.03 \\
\Lp\ & 10.52$\pm$0.06  & 11.16$\pm$0.06 & 11.39$\pm$0.06 \\
\enddata
\tablenotetext{a}{calculated from composite photometry using $\Delta$mag from Table \ref{nirc2}}
\tablenotetext{b}{from the 2MASS All-Sky Point Source Catalog \citep{cutri03}}
\label{binphot}
\end{deluxetable}

\begin{deluxetable}{lc}
\tablecolumns{2}
\footnotesize
\tablecaption{Properties of \sciencebin}
\tablewidth{0pt}
\tablehead{
}
\startdata
Angular Separation\tablenotemark{a}                                & 0.1582$\pm$0.0003\arcsec \\
Position Angle\tablenotemark{a}                         & 99.52$\pm$0.09$^\circ$ \\   
Distance & 325$^{+72}_{-50}$~pc \\
Projected Separation\tablenotemark{b} & 51$^{+11}_{-8}$~AU \\
Orbital Period & 1000$^{+1600}_{-500}$~yr \\
Proper Motion (mas yr$^{-1}$)             		& $-$1$\pm$6, 3$\pm$5\\
Radial Velocity\tablenotemark{c} (km s$^{-1}$)  & $-$10.6$\pm$0.5 \\
\enddata
\tablenotetext{a}{values measured from the $J$-band image.}
\tablenotetext{b}{calculated for a distance of 325$^{+72}_{-50}$~pc}
\tablenotetext{c}{the weighted mean of values in Table \ref{hires}}
\label{binprops}
\end{deluxetable}

\begin{deluxetable}{lrrrr}
\tablecolumns{5}
\footnotesize
\tablecaption{HIRES measurements of \sciencebin}
\tablewidth{0pt}
\tablehead{
\colhead{Date (UT)}			&
\colhead{Li EW}                  &
\colhead{H$\alpha$ EW}                  &
\colhead{H$\alpha$ 10\% width }             &
\colhead{Radial Velocity}  \\
\colhead{}			&
\colhead{\AA}                  &
\colhead{\AA}                  &
\colhead{km s$^{-1}$}             &
\colhead{km s$^{-1}$}             \\
}
\startdata
2006 May 11	&0.81$\pm$0.03 & $-$52.7$\pm$0.1	& 314$\pm$4 &$-$10.37$\pm$0.52\\
2006 Aug 12	&0.84$\pm$0.03 & $-$39.2$\pm$0.1	& 298$\pm$4 &$-$11.36$\pm$0.95\\
\enddata
\label{hires}
\end{deluxetable}

\end{document}